\documentclass[prd,amsmath,amssymb,superscriptaddress,preprintnumbers,twocolumn,nofootinbib,10pt]{revtex4-1}
\usepackage{amsmath,amssymb,graphics,epsfig,subfigure}
\usepackage{color}
\usepackage{mathrsfs}

\pdfoutput=1
\usepackage{graphicx}
\usepackage{dcolumn}
\usepackage{bm}
\usepackage{amssymb}
\usepackage{latexsym}
\usepackage{booktabs}
\usepackage{amsmath}
\usepackage{multirow}
\usepackage{url}
\usepackage{footnote}
\usepackage{float}
\usepackage{threeparttable}
\usepackage[colorlinks=true, linkcolor=blue, citecolor=blue]{hyperref}
\usepackage[bottom]{footmisc}

\usepackage[normalem]{ulem}
\usepackage{color}
\usepackage{array}
\usepackage{enumerate}
\usepackage{adjustbox}

\usepackage{makecell}
\usepackage{diagbox}
\usepackage{epstopdf}
\usepackage{epsfig}
\usepackage{longtable}
\usepackage{supertabular}
\usepackage{algorithm}
\usepackage{pifont}
\usepackage{algorithmic}
\usepackage{changepage}
\usepackage{setspace}

\usepackage{orcidlink}
\usepackage{etoolbox}

\begin{document}

	\title{Phase transitions in scalarized topological AdS black holes}

	\author{Zi-Qiang Zhao\orcidlink{0009-0009-7859-3655}}%\email{zhaoziqiang@stumail.neu.edu.cn}
	\affiliation{Liaoning Key Laboratory of Cosmology and Astrophysics, College of Sciences, Northeastern University, Shenyang 110819, China}

	\author{Zhang-Yu Nie\orcidlink{0000-0001-7064-247X}}%\thanks{Corresponding author}
	\email{niezy@kust.edu.cn}
	\affiliation{Center for Gravitation and Astrophysics, Kunming University of Science and Technology, Kunming 650500, China}
	
	\author{Shao-Wen Wei\orcidlink{0000-0003-0731-0610}}%\thanks{Corresponding author}
	\email{weishw@lzu.edu.cn}
	\affiliation{Lanzhou Center for Theoretical Physics, Key Laboratory of Theoretical Physics (Gansu) \& Key Laboratory of Quantum Theory and Applications (MOE), Lanzhou University, Lanzhou 730000, China}
	
	%\affiliation{Center for Gravitational Wave Experiment, Institute of Mechanics, Chinese Academy of Sciences, Beijing 100190, China}%
	
	\author{Jing-Fei Zhang\orcidlink{0000-0002-3512-2804}}%\email{jfzhang@mail.neu.edu.cn}
	\affiliation{Liaoning Key Laboratory of Cosmology and Astrophysics, College of Sciences, Northeastern University, Shenyang 110819, China}
	
	\author{Xin Zhang\orcidlink{0000-0002-6029-1933}}\email{zhangxin@neu.edu.cn}
	\affiliation{Liaoning Key Laboratory of Cosmology and Astrophysics, College of Sciences, Northeastern University, Shenyang 110819, China}
	\affiliation{MOE Key Laboratory of Data Analytics and Optimization for Smart Industry, Northeastern University, Shenyang 110819, China}
	\affiliation{National Frontiers Science Center for Industrial Intelligence and Systems Optimization, Northeastern University, Shenyang 110819, China}

	\begin{abstract}
		%我们研究了带电标量场模型在三种不同拓扑下的黑洞标量化过程的普适行为，结果表明，它们在低温下都会转变成标量化的黑洞。特别的，球拓扑更加特殊，它的标量化区间在低压强下，理论上可以拓展到非常高的温度。此外，我们还发现球拓扑下的标量化过程不需要非线性项就可以展现出类似于平面和双曲拓扑的复杂的相变行为。具体来讲，它们都存在一个相同的趋势，随着压强的增加，它们会从一阶相变转变为cave of wind相变，然后进入超临界区域。
		
		%这个研究可以更好的帮助我们理解黑洞标量化过程中的零阶相变问题，以及揭示黑洞在拓展相空间中的完整的相结构。
		
		We investigate the behavior of black hole scalarization induced by a charged scalar field in the extended phase space of the asymptotic AdS spacetime with three distinct horizon topologies. The results indicate that in all three cases, the charged black hole spacetime undergoes scalarization at low temperatures. Notably, the spherical topology is unique in that its domain of scalarization theoretically extends to much higher temperatures under low pressure in the extended phase space. Moreover, the scalarization process in the spherical case exhibits complex phase transition behaviors without additional non-linear terms, which are similar to those in the planar and hyperbolic topologies with the assistance of non-linear terms. With increasing pressure in the extended phase space, the condensate of the scalarization in all three cases undergoes a transition from the first-order style to a cave-of-wind style. This study provides deeper insight into the zeroth-order phase transition during black hole scalarization and reveals the complete phase structure of black holes in the extended phase space.
		
	\end{abstract}
	
	\keywords{Classical black hole, thermodynamics, phase transition}

	\pacs{04.70.Dy, 04.70.Bw, 05.70.Ce}

	\maketitle
	\section{Introduction}
	Recent studies in gravitational theories have uncovered a remarkable phenomenon, spontaneous scalarization, which poses a novel challenge to the extensions of the no-hair theorem \cite{Bartnik:1988am,Herdeiro:2015waa}. Spontaneous scalarization is a phase transition in which compact objects develop a nontrivial scalar ``hair'' due to a tachyon instability. Initially identified in scalar-tensor theories for neutron stars~\cite{Damour:1993hw}, this phenomenon has since been extended to black holes. In extended scalar-tensor-Gauss–Bonnet (ESTGB) gravity~\cite{Silva:2017uqg,Doneva:2017bvd,Antoniou:2017acq}, Schwarzschild black holes can undergo spontaneous scalarization, with curvature serving as the triggering mechanism. A similar effect occurs in Reissner–Nordström (RN) black holes within Einstein–Maxwell–scalar (EMS) models~\cite{Herdeiro:2018wub}. In certain ESTGB theories, Kerr black holes exhibit spin-induced scalarization~\cite{Cunha:2019dwb}. Furthermore, a considerable number of analogous studies on scalarization have been conducted \cite{Herdeiro:2020wei,Berti:2020kgk,Garcia-Saenz:2021uyv,Zhang:2022cmu}. These scalarization mechanisms share a common feature: the general hairless solutions remain valid but become unstable in certain regions of the parameter space, bifurcating into new families of scalarized black holes that are thermodynamically preferred. These hairy black holes generally exhibit richer phenomena.
	%近年来，引力理论的研究发现了一类引人注目的现象——自发标量化，它对无毛定理提出了新的挑战。自发标量过程是一种相变，在此过程中致密天体由于快子不稳定性而发展出非平凡的标量"毛"。该现象最初在标量-张量理论中发现于中子星，后来被扩展到黑洞。在扩展标量-张量-高斯-博内（ESTGB）引力中，史瓦西黑洞可以自发标量化，其中曲率提供了触发机制 。类似的现象，即电荷诱导的标量化，也发生在爱因斯坦-麦克斯韦-标量（EMS）模型的雷斯纳-诺斯特朗姆黑洞中。在某些ESGB理论中，克尔黑洞演示了自旋诱导的标量化。这些标量化机制共享一个共同的基本原理：广义相对论解仍然是可行的，但在参数空间的某些区域会变得不稳定，并分岔出新的、在动力学上更优的标量化黑洞族。这些带毛黑洞通常展现出更丰富的现象学。
	%除此之外，还有非常多类似的标量化的工作。
	
	The rich phase transition behavior of scalarized black holes in asymptotic AdS spacetime also provides a fruitful framework for investigating the AdS/CFT correspondence \cite{Maldacena:1997re,Hartnoll:2008vx,Hartnoll:2008kx,Cai:2010cv,Cai:2010zm,Li:2011xja,Cai:2013aca,Cai:2015cya}. In a broader context, the cosmological constant can be treated as thermodynamic pressure within the extended phase space, leading to rich phase structures \cite{Kubiznak:2012wp,Gunasekaran:2012dq,Cai:2013qga,Wei:2015iwa,Wei:2022dzw,Zhao:2025ecg}. Scalarization in the extended phase space of the asymptotic AdS spacetime reveals the formation of scalarized black holes below a certain critical temperature, as well as the zeroth order phase transition at a lower temperature~\cite{Guo:2021zed,Guo:2021ere}, which indicates the global instability of the system~\cite{Zhao:2022jvs}. In a model with the positive kinetic energy term and the potential term bounded from below, the system should be globally stable, indicating that new solutions with lower thermodynamic potential are waiting to be discovered. This leads to two fundamental questions: Do these new stable solutions really exist, and what is the final phase diagram in the extended phase space of the black holes?
	%How do we understand this instability taking place in the extended phase space of black holes? 
	%这引出了一个问题，我们应该怎么去理解这种拓扑相空间中的不稳定的相变模型？

	In this paper, we investigate the spontaneous scalarization of topological AdS black holes, revealing a phenomenon across different horizon geometries (spherical, planar, and hyperbolic). We consider a model where a charged scalar field with non-linear self-interactions couples to gravity. We demonstrate that scalarization occurs at low temperatures in all three topologies, with the spherical case exhibiting particularly intricate behavior—including cave-of-wind (COW) phase transitions and critical phenomena even in the absence of non-linear terms. Furthermore, we show that the pressure in the extended phase space drives the system from a zeroth-order to a first-order phase transition and eventually to a supercritical region, in a manner analogous to the back-reaction parameter or non-linear term. A striking implication of our analysis is that, in the low-pressure regime, all spherical AdS black holes may undergo a first-order transition into a scalarized state. 
	
	The remainder of this paper is organized as follows. In Sec.~\ref{sec2}, we introduce the model adopted in this work. In Sec.~\ref{sec3}, we present the phase diagram of scalarized black holes. In Sec.~\ref{sec4}, we discuss the most special case of scalarization under spherical topology. Finally, we provide a summary in Sec.~\ref{sec5}.

	\section{Model}\label{sec2}
	There are various methods to induce spontaneous scalarization of black holes. In this paper, we employ a charged scalar field model that includes one non-linear term.
	This model is also used in the study of holographic superconductors~\cite{Hartnoll:2008vx,Hartnoll:2008kx,Zhang:2021vwp,Zhao:2022jvs,Zhao:2025tqq}. The non-linear term is introduced to generate zeroth-order phase transitions in planar and hyperbolic topologies, facilitating the study of the phenomenon. Spherical topology is special; it exhibits a zeroth-order phase transition without any higher-order non-linear term, a phenomenon also observed in an uncharged scalar field~\cite{Guo:2021zed}. The total action is
	\begin{align}
		&S=\,S_{M}+S_{G},\quad
		S_G=\,\frac{1}{2\kappa_g ^2}\int d^{4}x\sqrt{-g}\left(R-2\Lambda\right),\nonumber\\
		&S_M=\,\frac{1}{q^2}\int d^{4}x\sqrt{-g}\Big(-\frac{1}{4}F_{\mu\nu}F^{\mu\nu}
		-D_{\mu}\psi^{\ast}D^{\mu}\psi 
		\nonumber\\
		&~~~~~~~~~~~-\frac{m^{2}}{L^2}\psi^{\ast}\psi-\lambda(\psi^{\ast}\psi)^{2}
		\Big).\label{actionAll}
	\end{align}
	Here, $F_{\mu\nu}=\nabla_{\mu}A_{\nu}-\nabla_{\nu}A_{\mu}$ is the Maxwell field strength, and $D_{\mu}\psi=\nabla_{\mu}\psi-i A_\mu\psi$ is the standard covariant derivative term of the charged scalar field $\psi$. In which $\Lambda=-3/L^2$. Here, $L$ is the AdS radius, and in the black hole extended space, $L$ should be considered as the pressure of the black hole with $P=3/(8\pi L^2)$.
	
	%通过对标量场和矢量场做变分，我们可以得到物质场的运动方程。
	The equations of motion for the matter fields are derived by varying the action with respect to the scalar and electromagnetic fields. We use the following ansatz
	\begin{align}
		\psi=\psi(r),~A_\mu dx^\mu=\phi(r)dt.\label{anstaz}
	\end{align}
	We will discuss the phenomenon we have discovered separately under spherical, planar, and hyperbolic topologies. The metric is as follows
	\begin{align}
		&ds^{2}=-N(r)\sigma(r)^2dt^{2}+\frac{1}{N(r)}dr^{2}+r^{2}d\Sigma^2_2,\label{MetricflatBR}
	\end{align}
	% Here
	% \begin{align}
		% d\Sigma^2_2 = \left\{
		% \begin{array}{ll}
			%     d\theta^2+sin^2\theta d\varphi^2 & \text{for } k=1, \\
			%     dx^{2}+dy^2 & \text{for } k=0,  \\
			%     d\theta^2+sinh^2\theta d\varphi^2 & \text{for } k=-1,
			% \end{array}
		% \right.
		% \end{align}
	in which
	\begin{align}
		N(r) =\frac{r^2}{L^2}-\frac{2M(r)}{r}+k,\label{sphereM}
	\end{align}
	where $k=$ 1, 0, and -1 for spherical, planar, and hyperbolic topologies, respectively. The Einstein field equation is
	\begin{align}
		R_{\mu\nu}-\frac{1}{2}(R-2\Lambda)g_{\mu\nu}=b^2\mathcal{T}_{\mu\nu},\label{Einst}
	\end{align}
	where $b=\kappa_g/q$ is the strength of the back-reaction of matter fields on the background geometry and $\kappa_g^2=8\pi G$. $\mathcal{T}_{\mu\nu}$ is the stress-energy tensor of the matter fields
	\begin{align}
		\mathcal{T}_{\mu\nu}=&(-\frac{1}{4}F_{\alpha\beta}F^{\alpha\beta}
		-D_{\alpha}\psi^{\ast}D^{\alpha}\psi-\frac{m^{2}}{L^2}\psi^{\ast}\psi-\nonumber\\
		&\lambda(\psi^{\ast}\psi)^{2})g_{\mu\nu}+(D_{\mu}\psi^{\ast}D_{\nu}\psi+D_{\nu}\psi^{\ast}D_{\mu}\psi)+\nonumber\\
		&F_{\mu\alpha}F^{\alpha}_{\nu}.
	\end{align}
	%求解爱因斯坦场方程，我们就可以得到包含物质场和时空背景的完整场方程
	By solving the Einstein field equation, we can obtain the complete set of field equations encompassing both matter fields and the spacetime background. We present the complete equations of motion and boundary conditions in Appendix~\ref{app:equations}. The Hawking temperature of these black hole metrics is
	\begin{align}
		T=\frac{N'(r_h)\sigma(r_h)}{4 \pi},
	\end{align}
	where $r=r_h$ labels the radius of the event horizon. By evaluating the on-shell action, we obtain the black hole's free energy. In this work, we consistently work within the canonical ensemble framework, which means fixing the total charge $\rho$. The Gibbs free energy equals the temperature times the Euclidean on-shell action in the bulk spacetime which reads
	\begin{align}
		\widetilde{G}=-\frac{1}{\beta}ln\mathcal{Z}, \mathcal{Z}=e^{-S^E},
	\end{align}
	in which
	\begin{align}
		S^E=&-\frac{1}{2\kappa_g^2}\int d^{4}x\sqrt{-g}(R-2\Lambda+2b^2\mathcal{L}_{M})\nonumber\\
		&+\frac{1}{\kappa_g^2}\int d^{3}x\sqrt{-h}\left(b^2 n_a F^{ab}A_b+K+\frac{2}{L}+\frac{2k}{r^2}\frac{L}{2}\right),
	\end{align}
	where $K$ is the trace of the extrinsic curvature $K_{\mu\nu}$ for the boundary (see, \textit{e.g.}, Refs.~\cite{Cai:2002mr,Nie:2014qma}). Here $K_{\mu\nu}=-h_\mu^{~\rho}\nabla_\rho n_\nu$ and $n$ is the unit normal to the boundary surface. For simplicity and without loss of generality, we omit the spacetime integral constant, yielding the final form of the free energy as
	\begin{align}
		G_k=\frac{2\kappa_g^2\widetilde{G_k}}{V_2}.
	\end{align}
	Here, $V_2$ is just the volume of the spatial boundary manifold. The complete free energy formula is presented in Appendix \ref{app:FEfunction}.
	%完整的自由能计算公式展示在了附录B中
	
	\section{Phase structure}\label{sec3}
	%对于平直AdS时空来说，加上标量场之后，系统在低于临界温度Tc之后就会发生自发标量化，即对称性自发破缺，此时黑洞系统就带上了额外的标量毛，并且此时系统的自由能更低。但是当我们加上一个负的高阶非线性项之后，这一支标量解就不再是一个稳定的解，但他仍然存在一小段稳定区域，也就是所谓的零阶相变。这一小段稳定区域的大小取决于自耦和参数的大小。零阶相变随着反作用参数的增大会逐渐被修正，它会逐渐变成一阶相变，并且最终在某个临界点进入超临界标量黑洞区域。我们已经在文章（）中详细讨论过这件事情。令人惊讶的是，这种临界现象并非是反作用参数独有的。当我们把注意力放到压强上，我们会发现，这种临界现象依然存在。
	% For asymptotically flat AdS spacetime, the introduction of a scalar field leads to spontaneous scalarization of the system below a critical temperature $T_c$ resulting in spontaneous symmetry breaking.
	The black hole in an asymptotically planar AdS spacetime undergoes spontaneous scalarization when a charged scalar field is introduced. This occurs below a critical temperature $T_c$ and leads to spontaneous symmetry breaking. Consequently, the black hole system acquires additional scalar hair, and the system's free energy becomes thermodynamically favorable. However, when a negative higher-order non-linear term $\lambda(\psi^*\psi)^2$ is added, this scalarized solution is no longer stable; yet, it still exists in a small stable region known as the zeroth-order phase transition~\cite{Cai:2013aca,Guo:2021zed,Zhao:2022jvs}. The size of this stable region depends on the value of the self-interaction parameter $\lambda$. This zeroth-order phase transition is progressively modified as the back-reaction parameter increases, gradually transitioning into a first-order phase transition and ultimately entering the supercritical scalarized black hole region at a certain critical point. We have discussed these results in detail in Ref.~\cite{Zhao:2025tqq}. This critical phenomenon is not unique to the back-reaction parameter. When focusing on pressure, we find that this critical phenomenon also persists.

	%在图1中，我们给出了不同压强下平面拓扑的自由能与温度的关系，并且给出了相应的landscape分析。在这篇文章中，我们并没有精确计算landscape，而是按照自由能的参数给出了landscape的猜测图像，这意味着只有极值点是满足运动方程的。在图1中，红色实线为一阶相变，随着压强增加变成蓝色实线表示的COW相变，并且最终会变成品红色实线表示的超临界标量黑洞。我们在（）中分析过，零阶相变是一种不稳定的模型，只要它存在另外一个高阶非线性项，不管这个非线性项多小，零阶相变总会在标量场足够大的时候转变成一阶相变。（）则表明了反作用参数有着等同于高阶非线性项的效应。图1中给出的结果则表明，压强同样有着等同于高阶非线性项的效应，这意味着，只要压强不为零，所有的零阶相变都会在温度足够大的时候转变为一阶相变。
	%不同的颜色的曲线对应不同的压强。圆圈代表稳态，正方形代表亚稳态，倒三角代表不稳定态。
	%%%%%%%%%%%%%%%%%%%%%%%%%%
	\begin{figure}[t]
		\center
		\includegraphics[width=\columnwidth]{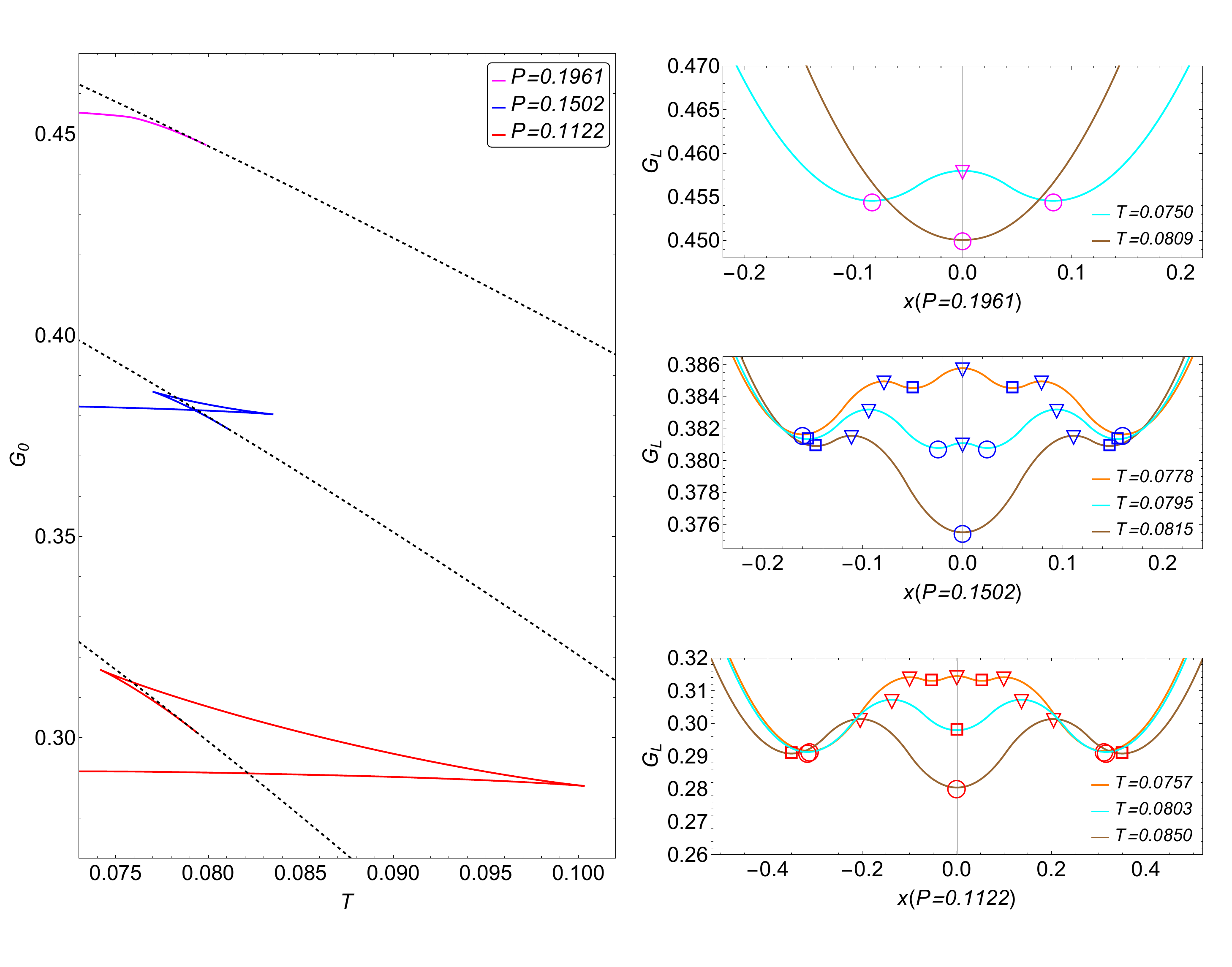}
		\caption{The free energy and landscape.
			Left: The relationship between free energy and temperature for plane topology under different pressures, where the red solid line indicates a first-order phase transition, blue solid lines represent COW phase transitions, magenta solid lines denote second-order phase transitions, and black dashed lines are the normal solution. Right: Landscape analysis corresponding to different pressures on the left-hand side. Different colors represent different pressures. Circles denote stable states, squares represent metastable states, and inverted triangles indicate unstable states. The value of $G_L$ corresponds to that of the conjectured landscape analysis, where only the extremal points satisfy the equations of motion.
		}\label{landscapePlaneP}
	\end{figure}
	%%%%%%%%%%%%%%%%%%%%%%%%%%
	%Gl对应于猜测的landscape分析的值，其中只有极值点满足运动方程
	
	%COW相变主要用于区分传统的一阶相变（相变点发生在normal解和标量解之间），COW相变先发生了一次二阶相变，存在一个标量解1，然后通过一阶相变从标量解1转变到了标量解2.
	
	In Fig.~\ref{landscapePlaneP}, we show the free energy as a function of temperature for the planar topology at different pressures, with the corresponding landscape analysis. In this paper, we do not compute the landscape exactly but rather provide its conjectured form based on the values of the free energy, which means only the extremal points satisfy the equations of motion. 
	The landscape picture is primarily used to understand spontaneous symmetry breaking in phase transitions and the metastable region of first-order phase transitions. Within the thermodynamic framework, the landscape picture is a very useful tool. For static equilibrium solutions of first-order phase transitions, three typical equilibrium states exist: stable, metastable, and unstable. Among them, the unstable state corresponds to a saddle point in the landscape, which separates the local minima of the stable and metastable states. When nonequilibrium dynamical evolution is considered, this saddle point facilitates bubble formation  \cite{Janik:2017ykj,Li:2020ayr,Chen:2022tfy,Zhao:2023ffs,Zhao:2026eav,Jin:2026lzp,Zhao:2026als}. For equilibrium solutions, however, only the extrema of the landscape satisfy the equations of motion; other points must be obtained by evaluating the off-shell action. Since our paper only considers equilibrium physics, the landscape picture inferred in this way is sufficient for analyzing the physical processes under consideration.
	In Fig.~\ref{landscapePlaneP}, the red solid line corresponds to first-order phase transitions. With increasing pressure, it becomes the blue solid line indicating the COW phase transition, and eventually becomes the magenta solid line indicating supercritical scalarized black holes. The COW phase transition is primarily used to distinguish it from conventional first-order phase transitions. Unlike a typical first-order transition occurring between a normal black hole and a scalarized black hole, the COW phase transition involves an initial second-order phase transition to a scalarized black hole solution 1, followed by a first-order transition from scalarized solution 1 to scalarized solution 2~\cite{Zhao:2022jvs,Zhao:2025tqq}. The analysis in Ref.~\cite{Zhao:2022jvs} shows that zeroth-order phase transitions are unstable configurations, when there exists any additional higher-order non-linear term, no matter how small, the zeroth-order transition will always turn into a first-order phase transition when the scalar field becomes sufficiently large. Ref.~\cite{Zhao:2025tqq} further demonstrates that the back-reaction parameter $b$ has effects equivalent to higher-order non-linear terms. The results presented in Fig.~\ref{landscapePlaneP} indicate that pressure also exhibits effects similar to higher-order non-linear terms, meaning that as long as the pressure is non-zero, all zeroth-order phase transitions will transform into first-order phase transitions at sufficiently high temperatures.
	
	%我们在图2中分别给出了平面拓扑随着反作用参数和压强的变化情况的相图，其中红色实线表示COW相变中两个不同标量黑洞解的相变点，红色虚线表示普通黑洞和标量黑洞之间的一阶相变的相变点，蓝色虚线则表示一阶相变的斯宾诺区域。可以看到，他们最终都会到达一个临界点，这意味着黑洞此时已经变成了超临界标量黑洞。图2c则给出了双曲拓扑下压强与温度的相图，它展现出来的相变行为和平面拓扑下的行为是一致的。
	%%%%%%%%%%%%%%%%%%%%%%%%%%%
	\begin{figure*}[!htbp]
		\subfigure[]{\label{flatFixB}\includegraphics[width=4.8cm]{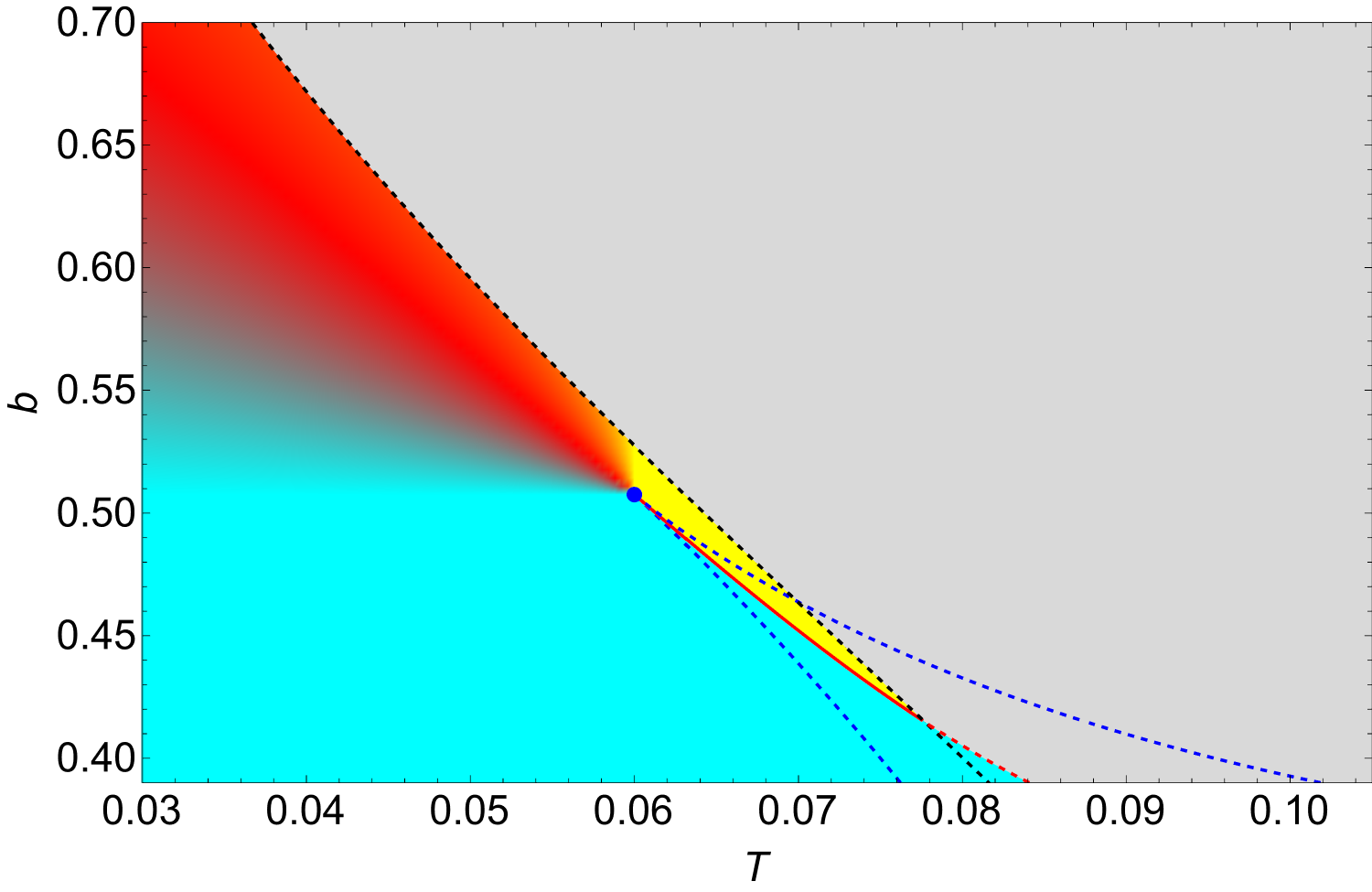}}
		\subfigure[]{\label{flatFixP}\includegraphics[width=5cm]{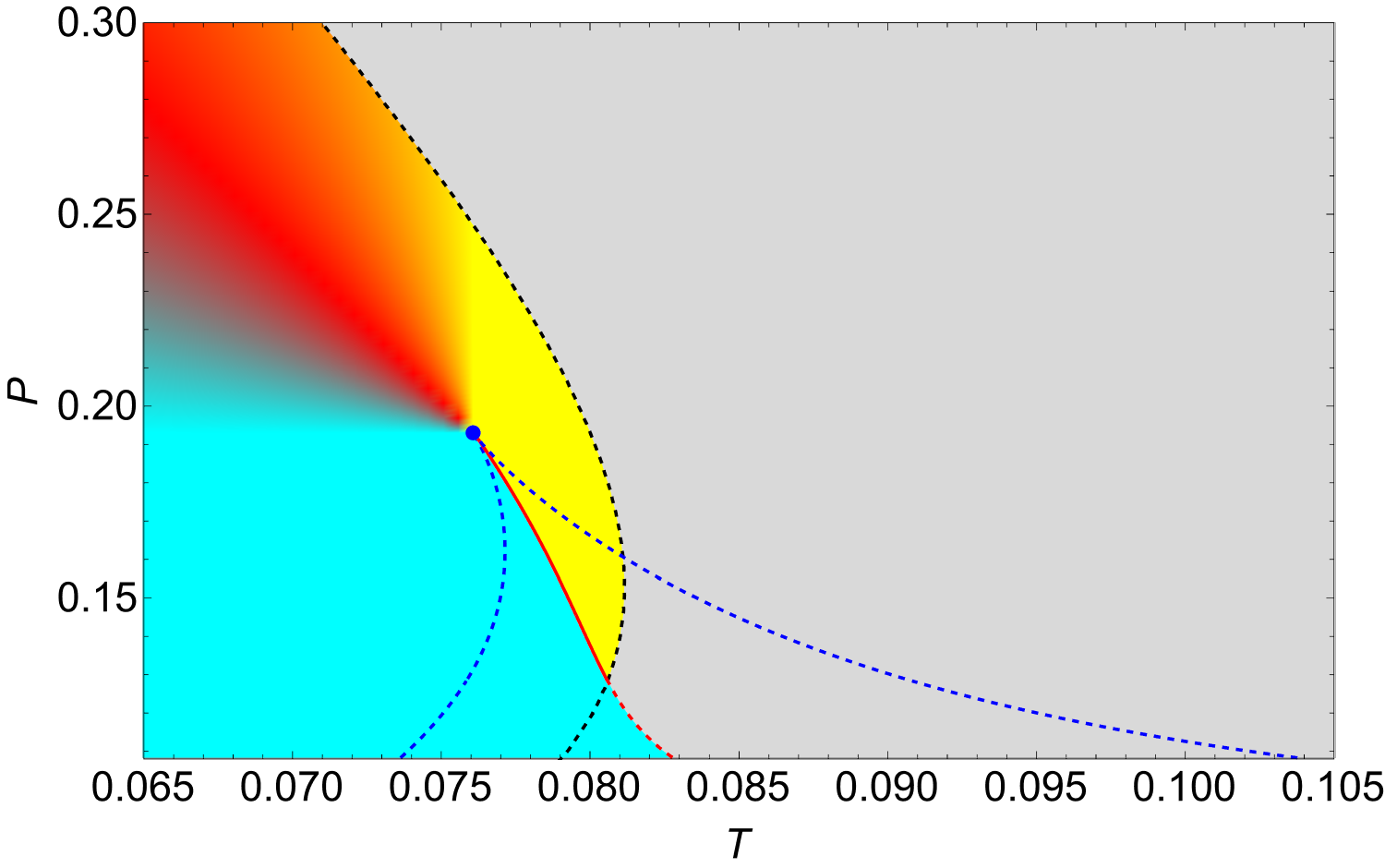}}
		\subfigure[]{\label{hyperbolicFixP}\includegraphics[width=5cm]{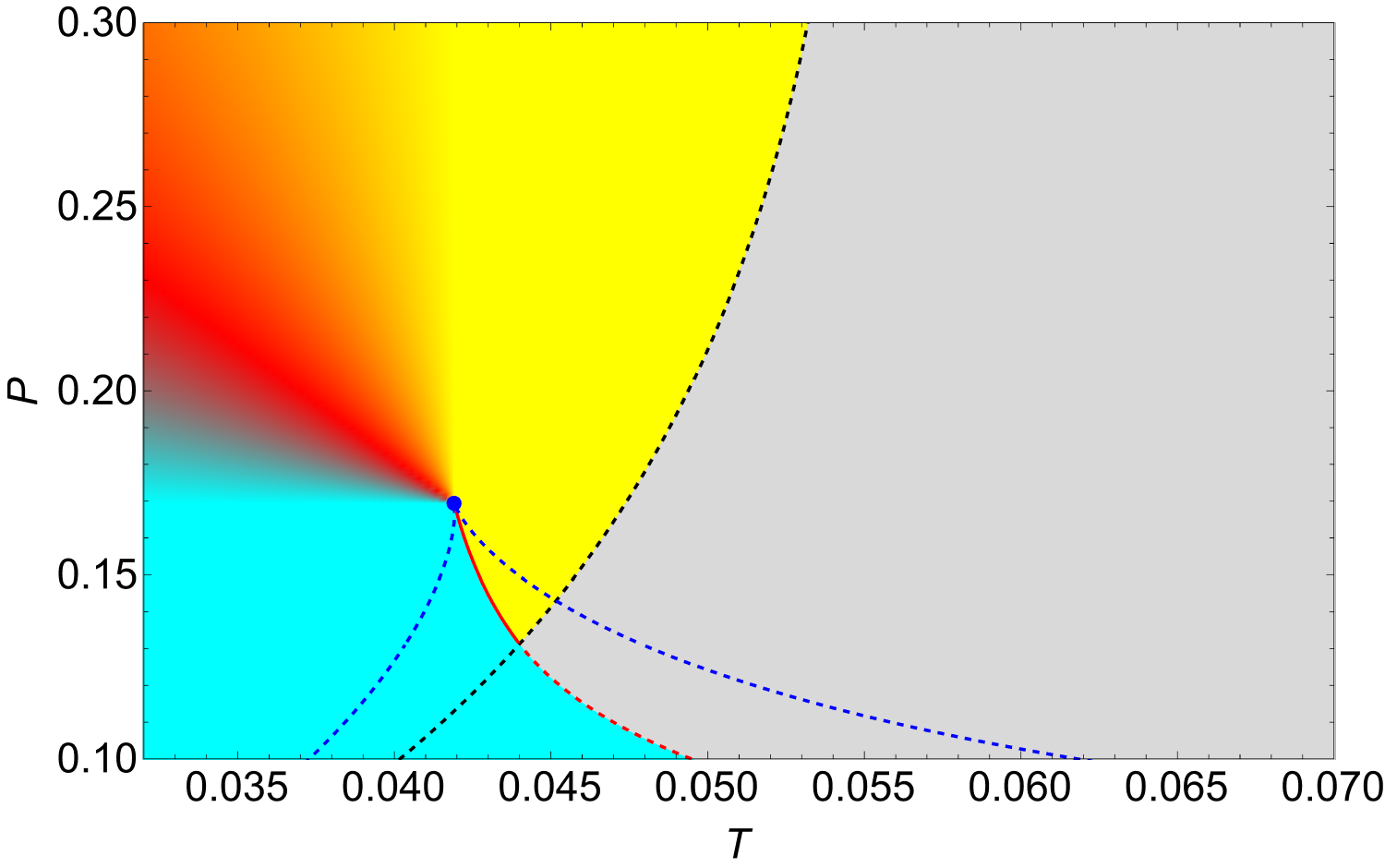}}
		\caption{
			The phase diagrams of planar and hyperbolic topology. Panel (a): The phase diagram of $b-T$ for planar topology with fixed $L=1$ and $\lambda=-0.35$. Panel (b): The phase diagram of $P-T$ for planar topology with fixed $b=0.4$ and $\lambda=-0.35$. Panel (c): The phase diagram of $P-T$ for hyperbolic topology with fixed $b=0.4$ and $\lambda=-0.2$. The red solid and red dashed lines represent the COW and first-order phase transitions. The blue dashed line delineates the spinodal regions for these phase transitions.
			The gray area signifies the normal solution, while the yellow and cyan regions correspond to two distinct hairy scalar solutions. The red region denotes the supercritical regime.
		}\label{phaseDiagram_flat_hpyerbolic}
	\end{figure*}
	%%%%%%%%%%%%%%%%%%%%%%%%%%%

	%%%%%%%%%%%%%%%%%%%%%%%%%%%
	\begin{figure*}[!htbp]
		\subfigure[]{\label{fixdP}\includegraphics[width=5cm]{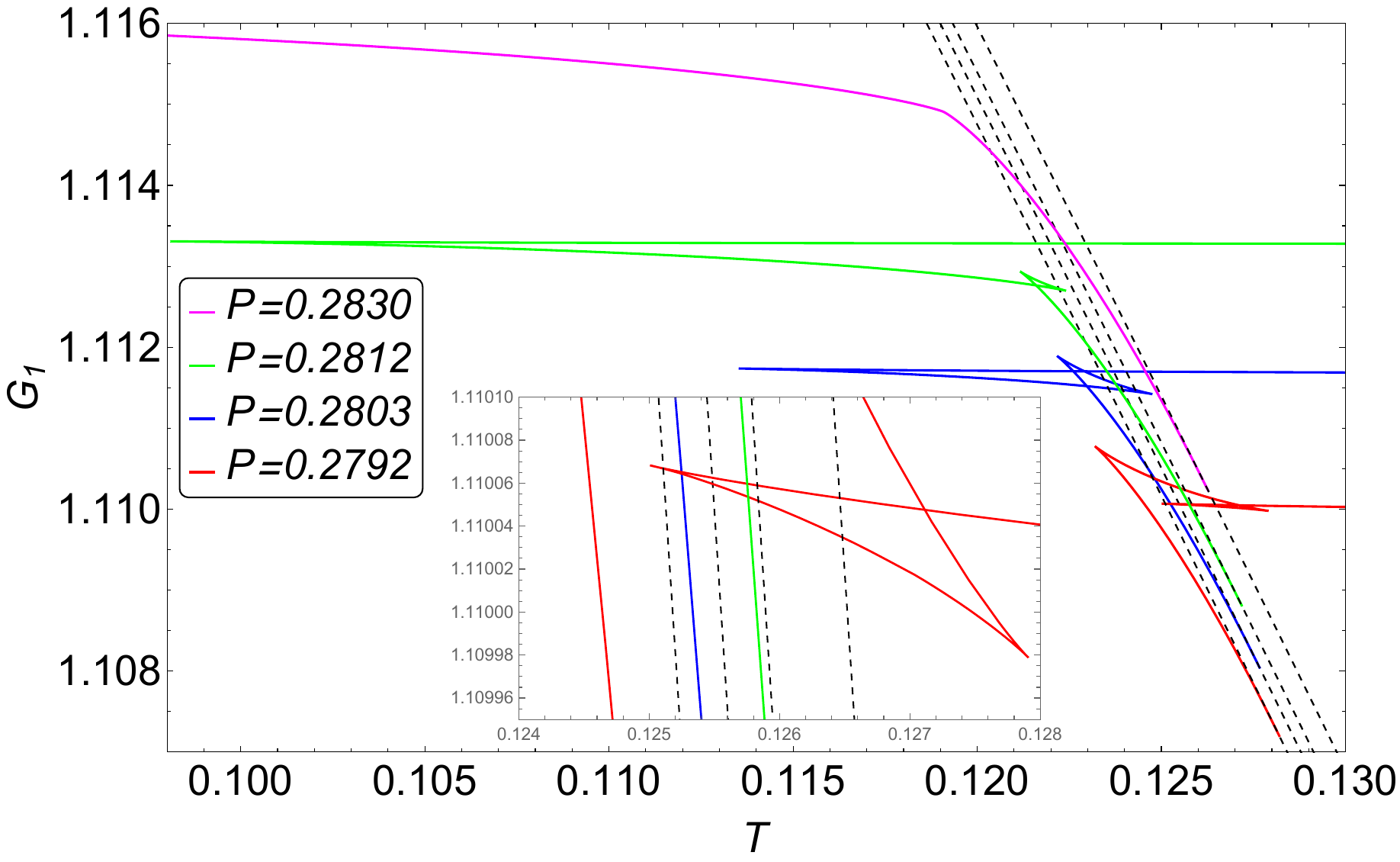}}
		\subfigure[]{\label{phaseDiagram_large}\includegraphics[width=5cm]{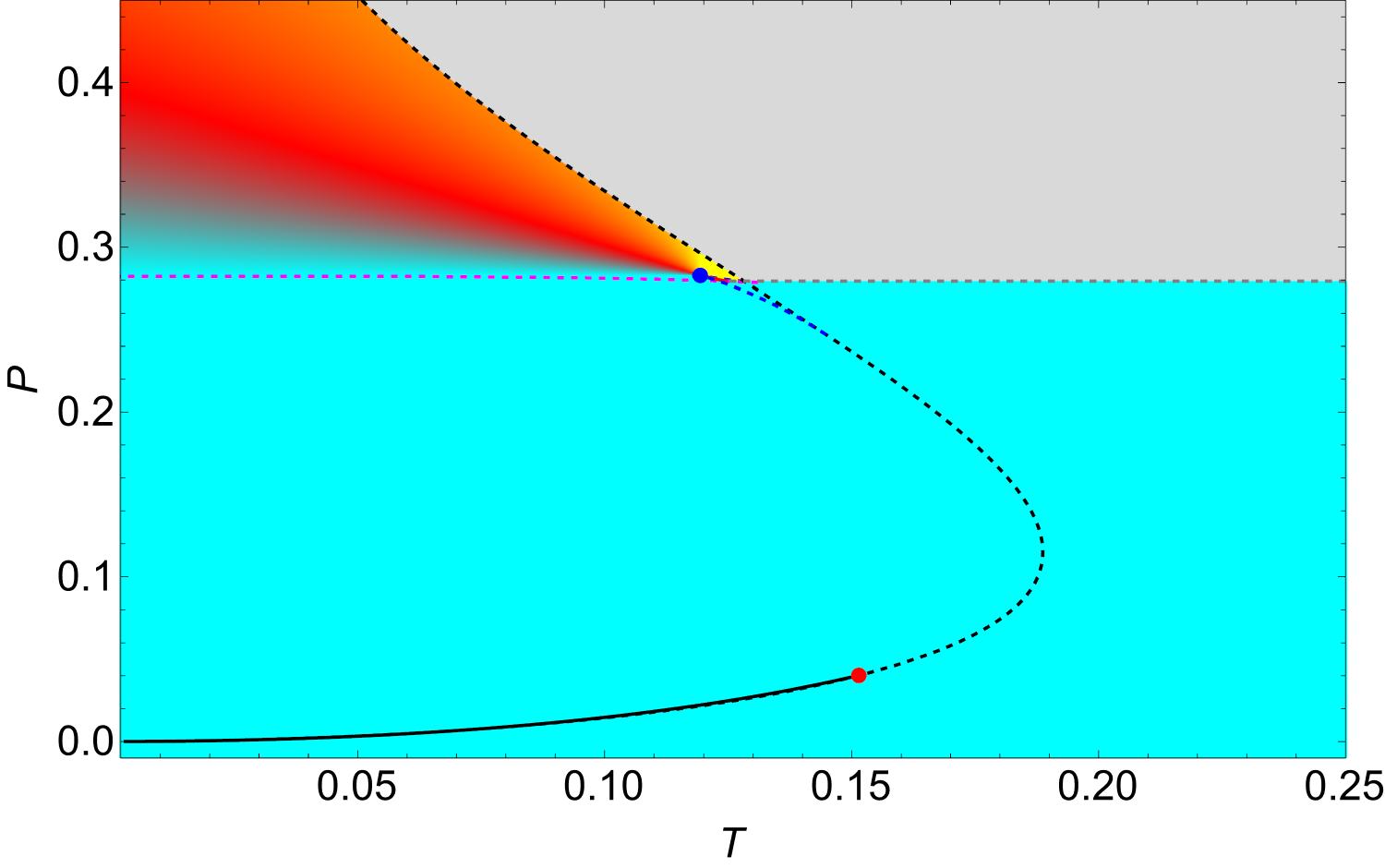}}
		\subfigure[]{\label{phaseDiagram_small}\includegraphics[width=5cm]{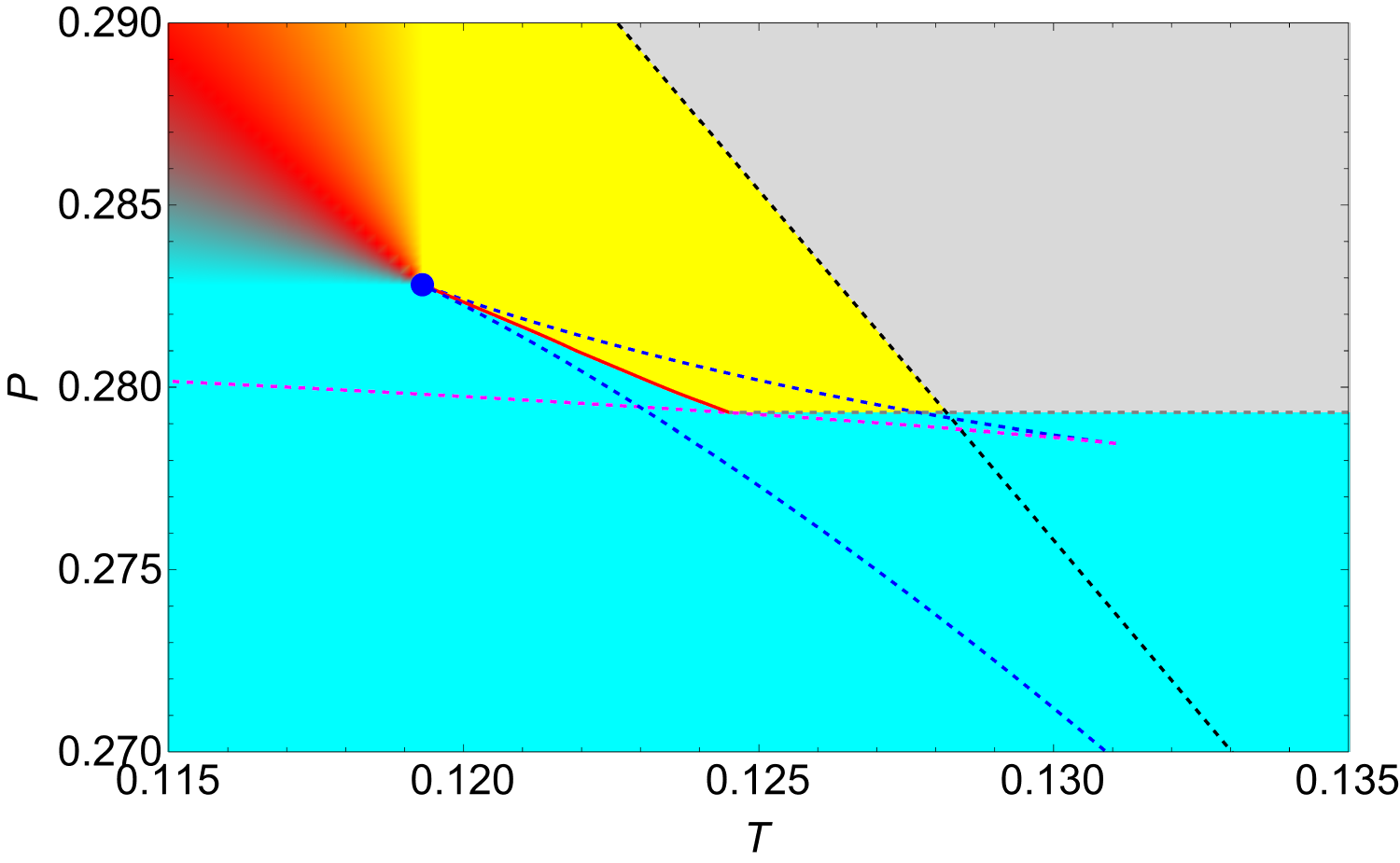}}
		\caption{
			%lambda=0tau=0b=0.4时候，度规3的自由能和相图。
			The free energy and phase diagram for sphere topology with $\lambda=0$ and $b=0.4$. Panel (a): The black dashed and solid lines are for the normal and scalarized solutions. Panels (b) and (c): The phase diagram of sphere topology. The red region is the supercritical scalarized black hole region, the yellow and cyan regions are the two different scalarized solution regions. Panel (c): The enlarged diagram of (b).
		}\label{RNadsPhase}
	\end{figure*}
	%%%%%%%%%%%%%%%%%%%%%%%%%%%

	Figs.~\ref{flatFixB} and \ref{flatFixP} show phase diagrams for planar topology under varying back-reaction and pressure, respectively. The red solid line marks transition points between different scalarized black hole solutions in COW phase transitions. 
	The red dashed line indicates the transition points of first-order phase transitions between normal and scalarized black holes. The blue dashed lines show the spinodal region of the COW and first-order phase transitions. All COW phase transitions terminate at a critical point before entering the supercritical regime. Fig. \ref{hyperbolicFixP} presents the $P-T$ phase diagram for the hyperbolic topology, exhibiting the same transition behavior as the planar cases.

	%s
	\section{Scalarization in spherical topology}\label{sec4}
	%现在让我们集中注意力到最有趣的情况，球拓扑。当作用量中带电标量场项psi等于0时，度规（）中k=1的情形将会回归到我们最熟悉的rn-ads黑洞。
	Let us now focus on the most interesting case: spherical topology. When $\psi=0$ is in the action (\ref{actionAll}), the metric (\ref{sphereM}) with $k=1$ reduces to the familiar RN-AdS black hole. The only difference is that in our model, the metric of the RN-AdS black hole will have an additional back-reaction parameter $b$
	\begin{align}
		N(r)&=1+\frac{b^2Q^2}{2r^2}+\frac{r^2}{L^2}-\frac{2M}{r},
	\end{align}
	where the mass can be expressed in terms of the event horizon radius
	\begin{align}
		M&=\frac{b^2Q^2}{4r_h}+\frac{r_h}{2}+\frac{r^3_h}{2L^2}.
	\end{align}
	In this subsection, we set $\lambda=0$, implying that the system lacks non-linear terms. For computational convenience, we choose $b=0.4$. We have also computed results for other values of $b$, with qualitatively similar behavior observed. At low pressure, the system exhibits the familiar first-order phase transition in the extended phase space of the RN-AdS black hole, eventually reaching a critical point with increasing pressure, beyond which the system enters the supercritical regime. This result corresponds to the black solid line in Fig.~\ref{phaseDiagram_large}, and the red point is the critical point of the RN-AdS black hole.
	%这个结果就是我们在图3中展示的红色实线
	
	%数值结果表明，此时黑洞系统存在一段自由能更高的非稳态标量黑洞解。如果我们继续增大压强，就会发现这一支非稳态标量解会出现一小段稳定区域，也就是我们熟悉的零阶相变，和我们在平直以及双曲拓扑中介绍过的增加一个非线性项lambda的结果一样。这种不稳定的相变模型理论上不应该存在，一个有效的做法是增加一个额外的高阶非线性项。但是正如我们在文章（）以及上一小节介绍过的，压强和反作用参数都拥有着和高阶非线性项相同的效应。所以，如果我们继续增大压强就会发现，这种不稳定的零阶相变会转化成一阶相变。
	
	Numerical results indicate the existence of an unstable scalarized black hole solution with higher free energy for the black hole system at this stage. If we continue to increase pressure, we find that this unstable solution transitions into a small stable region, corresponding to the zeroth-order phase transition we are familiar with from planar and hyperbolic topologies. Theoretically, such an unstable phase transition model should not exist. An effective approach is to introduce an additional higher-order non-linear term. However, as we discussed earlier in this paper and in the previous study \cite{Zhao:2025tqq}, both pressure and back-reaction parameters have similar effects to those of higher-order non-linear terms. Consequently, upon further increasing pressure, this unstable zeroth-order phase transition evolves into a first-order phase transition.
	
	%在图（）中，我们给出了标量黑洞随着压强的变化趋势。随着压强的增加，临界相变的非稳态会出现一小段态解，然后该稳态解会变成与亚稳态之间的一阶相变，也就是非稳态的COW相变。随着压强的继续增大，零阶相变的非稳态会消失，系统会呈现出标准的一阶相变燕尾曲线，并且最终进入超临界区域。这种行为与平面和双曲拓扑的行为是一样的。
	In Fig.~\ref{fixdP}, we show the phase transition behavior of scalarized black holes with increasing pressure. A metastable solution appears briefly at the unstable solution of the zeroth-order phase transition and then evolves into a first-order phase transition between stable and metastable states. With a further increase in pressure, the unstable branch of the zeroth-order phase transition rapidly approaches zero temperature. The system then exhibits a standard first-order phase transition characterized by a swallowtail curve, eventually entering a supercritical region. This behavior is consistent with planar and hyperbolic topologies.
	By calculating the free energy profiles across all pressure values, we obtain the phase diagram for scalarized black holes with spherical topology. Fig. \ref{phaseDiagram_large} shows the complete phase diagram, while Fig. \ref{phaseDiagram_small} provides a magnified view. 
	
	%正如我们前面分析过的那样，
	
	As previously analyzed, the spherical topology exhibits the same phase structure as planar and hyperbolic cases without requiring an additional higher-order non-linear term. Furthermore, consistent with landscape analysis and our results for planar/hyperbolic topologies, all zeroth-order transition instabilities should convert to stable first-order transitions when $\psi$ becomes sufficiently large, provided both back-reaction and pressure remain non-zero. Unfortunately, our numerical calculations cannot reach such large $\psi$ values. We thus base these conclusions on results from planar and hyperbolic topologies. Nevertheless, our study also shows that scalarized black holes with a spherically topological horizon exhibit a rich phase structure even without adding higher-order nonlinear terms under high pressure.

	%可惜的是，我们的数值计算没办法计算到如此大的psi值，我们只能从平面和双曲拓扑的结果中总结出上述结论。尽管如此，我们的研究也展示了在高压强下，具有球拓扑视界的标量化黑洞在即使不添加高阶非线性项的情况下也展现出了丰富的相结构。

	%先介绍一下总体研究了什么，第二段讲一下增加压强的普适结果，第三段讲球拓扑低压下总可能经过一阶相变转变为带毛黑洞
	\section{Conclusions}\label{sec5}
	This study investigates the spontaneous scalarization of topological AdS black holes--spherical, planar, and hyperbolic--coupled with a charged scalar field, revealing a similar phenomenon that occurs simultaneously for all three topologies. We illustrated that scalarization, a phase transition in which black holes spontaneously develop scalar hair at low temperatures, is a common feature across all three topologies.
	
	%我们的研究发现，拓展相空间中的压强可以作为调控标量黑洞的相变的有效参数，增加压强将直接导致标量黑洞系统的一阶相变到达临界点，随后进入超临界区域。这一普适结果对三种拓扑的黑洞都适用。此外，平面和双曲拓扑需要高阶非线性项让系统出现不稳定分支解，从而出现零阶相变，而球面拓扑则不需要，它天然存在一支自由能更高的不稳定解。
	Our study reveals that in the extended phase space, pressure controls the phase transitions of scalarized black holes. Increasing the pressure drives the first-order phase transition of the system to a critical point, beyond which the system enters a supercritical region. Contrarily, when we lower the pressure, the scalarization of the black holes turns into first-order phase transitions with increasing phase transition temperature, which makes it more difficult to find the final stable section of solutions with a very large condensate value. This behavior applies to black holes with all three topologies. Furthermore, while planar and hyperbolic topologies require higher-order non-linear terms to generate unstable branches of solutions in the first-order phase transition, the spherical topology inherently exhibits such unstable branches of solutions, requiring no additional non-linear terms.
	
	%此外，零阶相变作为一个不能出现的病态模型，我们发现，它从landscape上分析应该还存在一段自由能更低的稳定解。我们从同一个模型出发，分别在三种不同拓扑下进行了研究。压强和物质场反作用都可以有效的看成是某种高阶的非线性自耦合作用，在凝聚值足够大时使得系统再次稳定。这使得我们可以得到一个非常有意思的结果，即第压强下，常规黑洞可以通过一阶相变转变为标量黑洞，并且这个转变温度非常高。因此，球拓扑展现出来的零阶相变并非是一种病态的模型，它是一个相变温度无限大的一阶相变的亚稳态分支。
	In the previous studies, the scalarization phase transition in the low pressure region of the spherical topology case is claimed to be zeroth-order, and the scalarized solutions are all unstable. However, the zeroth-order phase transition is a pathological signal that should not occur in a stable setup. From a landscape analysis and with the above phenomenon, we speculated that the zeroth-order phase transition occurs because the system initially follows an incomplete branch of unstable solutions before tunneling to the stable phase; therefore, the unstable section of solutions is expected to turn back to a new branch of stable solutions with lower free energy, while the zeroth-order phase transition exhibited in spherical topology will be replaced by a first-order phase transition into scalarized black holes at a very high temperature.

	In summary, our work investigates the phase transition behaviors of scalarized black holes from the perspective of different topologies. It establishes the role of pressure as a key control parameter in the extended phase space of black holes and specifically reveals the unique and rich physics of spherical topology black holes during scalarization. These findings not only deepen the understanding of gravitational nonlinearity and black hole thermodynamics but also provide new insights for studying phase transitions in strongly coupled systems within the AdS/CFT correspondence framework.

	\section*{Acknowledgements}
	We are grateful to Xin Zhao for useful discussions. This work is supported by the National Natural Science Foundation of China (grant nos. 12533001, 12575049, 12473001, 12205039, 12305058, 11965013 and 12575054). ZYN is partially supported by Yunnan High-level Talent Training Support Plan Young $\&$ Elite Talents Project (grant no. YNWR-QNBJ-2018-181). This work is also supported by the National SKA Program of China (grant nos. 2022SKA0110200 and 2022SKA0110203) and the 111 Project (grant no. B16009).
	
%\bibliography{reference}	
	
	% Appendix in single-column format
%	\newpage
	% 附录部分
	\onecolumngrid  % REVTeX 专用的单栏切换命令
	
	\appendix
	\section{Full Equations of Motion}
	\renewcommand{\theequation}{A-\arabic{equation}}
	\label{app:equations}
	\setcounter{equation}{0}
	%通过将度规和anstaz带入作用量，我们可以得到如下包含爱因斯坦场方程的完整的运动方程
	By substituting the metric (\ref{MetricflatBR}) and ansatz (\ref{anstaz}) into the action (\ref{actionAll}), we obtain the following complete set of equations of motion, including the Einstein field equations
	\begin{align}
		\psi''(r)=&\left(\frac{m^{2}}{L^2N(r)}-\frac{\phi(r)^{2}}{N(r)^2\sigma(r)}\right)\psi(r)-\left(\frac{N'(r)}{N(r)}+\frac{2}{r}+
		\frac{\sigma'(r)}{\sigma(r)}\right)\psi'(r)+\frac{1}{N(r)}2\lambda\psi(r)^3,\\
		\phi''(r)=&\frac{2\psi(r)^{2}}{N(r)}\phi(r)-\left(\frac{2}{r}-\frac{\sigma'(r)}{\sigma(r)}\right)\phi'(r),\\
		\sigma'(r)=&b^2r\left(\frac{\phi(r)^2\psi(r)^2}{N(r)^2\sigma(r)}+\sigma(r)\psi'(r)^2\right),\\
		M'(r)=&\frac{b^2r^2}{\sigma(r)^2}\left(\frac{\phi'(r)^2}{4}+\frac{\phi(r)^2\psi(r)^2}{2N(r)}\right)+\frac{1}{2}b^2r^2\left(N(r)\psi'(r)^2+\frac{m^2}{L^2}\psi(r)^2\right)+\frac{1}{2}b^2r^2\lambda\psi(r)^4.\label{EQofMotion}
	\end{align}
	To solve the above equations of motion, boundary conditions need to be imposed. The asymptotic expansion at the horizon takes the form
	\begin{align}
		\phi(r)=&\phi_{h_1}(r-r_h)+\phi_{h_2}(r-r_h)^2+\cdots,\\
		\psi(r)=&\psi_{h_0}+\psi_{h_1}(r-r_h)+\cdots,\\
		\sigma(r)=&\sigma_{h_0}+\sigma_{h_1}(r-r_h)+\cdots,\\
		M_1(r)=&\frac{r_h}{2}(1+\frac{r_h^2}{L^2})+M_{h_1}(r-r_h)+\cdots,\\
		M_0(r)=&\frac{r^3_h}{2}+M_{h_1}(r-r_h)+\cdots,\\
		M_{-1}(r)=&\frac{r_h}{2}(\frac{r_h^2}{L^2}-1)+M_{h_1}(r-r_h)+\cdots.
	\end{align}
	The asymptotic expansion at the AdS boundary takes the form
	\begin{align}
		\phi(r)&=\mu-\frac{\rho}{r}+\cdots,\\
		\psi(r)&=\frac{\psi^{(1)}}{r}+\frac{\psi^{(2)}}{r^2}+\cdots,\\
		\sigma(r)&=\sigma_{b_0}+\frac{\sigma_{b_3}}{r^3}+\cdots,\\
		M(r)&=M_{b_0}+\frac{M_{b_1}}{r}+\cdots.
	\end{align}
	For the electromagnetic field, in the absence of scalarization, its form is given by
	\begin{align}
		\phi(r)=\frac{\rho}{r_h}(1-\frac{r_h}{r}).
	\end{align}
	For computational simplicity, we implement the following coordinate transformation
	\begin{align}
		&\widetilde{\phi}(r)\rightarrow\frac{1}{r_h}\phi(\frac{r_h}{r}),~\widetilde{\psi}(r)\rightarrow\frac{r_h}{r}\psi(\frac{r_h}{r}),~\widetilde{M}(r)\rightarrow M(\frac{r_h}{r}),~\widetilde{\sigma}(r)\rightarrow\sigma(\frac{r_h}{r}).
	\end{align}
	This ensures the numerical solution range stays within $[0,1]$ while keeping the boundary conditions of $\phi(r)$ independent of $r_h$.
	The boundary conditions are $\widetilde{\sigma}(\infty)=1$, $\widetilde{\psi}^{(1)}(\infty)=0$, $\widetilde{\phi}(r_h)=0$, and $\widetilde{\phi}(\infty)=\rho=1$. We use the shooting method to solve the equations of motion.

	\section{Free energy and temperature}
	\renewcommand{\theequation}{B-\arabic{equation}}
	\label{app:FEfunction}
	\setcounter{equation}{0}
	The complete expression for the free energy is given below
	\begin{align}
		G_1&=2b^2\mu\rho-2M_{b0}+2r_h+2\int^{\infty}_{r_h}(1-\sigma(r))dr,\\
		G_0&=2b^2\mu\rho-2M_{b0},\\
		G_{-1}&=2b^2\mu\rho-2M_{b0}-2r_h+2\int^{\infty}_{r_h}(\sigma(r)-1)dr.
	\end{align}
	The complete temperature expression that incorporates the scalar field is presented below
	\begin{align}
		T_1=&\frac{3\sigma_{h0}r_h}{4 L^2\pi}-\frac{b^2\phi_{h1}^2}{8\pi \sigma_{h0}r_h^3}+\frac{\sigma_{h0}}{4\pi r_h}-\frac{b^2m^2\sigma_{h0}\psi_{h0}^2r_h}{4 L^2\pi},\\
		T_0=&\frac{3\sigma_{h0}r_h}{4 L^2\pi}-\frac{b^2\phi_{h1}^2}{8\pi \sigma_{h0}r_h^3}-\frac{b^2m^2\sigma_{h0}\psi_{h0}^2r_h}{4 L^2\pi}-\frac{b^2\lambda\sigma_{h0}\psi_{h0}^4r_h}{4 \pi},\\
		T_{-1}=&\frac{3\sigma_{h0}r_h}{4 L^2\pi}-\frac{b^2\phi_{h1}^2}{8\pi \sigma_{h0}r_h^3}-\frac{\sigma_{h0}}{4\pi r_h}-\frac{b^2m^2\sigma_{h0}\psi_{h0}^2r_h}{4 L^2\pi}-\frac{b^2\lambda\sigma_{h0}\psi_{h0}^4r_h}{4 \pi}.
	\end{align}
	
\twocolumngrid

\bibliography{reference}	
	
\end{document}